\documentstyle[pre,aps,multicol,epsf]{revtex} 
\begin{document}
\title{Freely decaying weak turbulence for sea surface 
gravity waves}

\author{M. Onorato, A. R. Osborne, M. Serio }
\address{Dip. di Fisica Generale, Universit\`{a} di Torino, Via
Pietro Giuria 1, 10125 Torino, Italy}
\author{D. Resio}
\address{Coastal and Hydraulics Laboratory, U.S. Army
 Engineer Research and Development Center, 
Halls Ferry Road, Vicksburg, MS 39180, USA}
\author{A. Pushkarev}
\address{Waves and Solitons LLC, 918 W. Windsong Dr., Phoenix, AZ 85045, USA}
\author{V. Zakharov}
\address{Landau Institute for Theoretical Physics, Moscow, 117334, Russia and
Department of Mathematics, University of Arizona, Tucson, AZ 85721, USA}
\author{C. Brandini}
\address{La.M.M.A., Regione Toscana, Via A. Einstein 35/b 
50013Ê Campi Bisenzio - Firenze, Italy}
\maketitle
\begin{abstract}
We study numerically the generation of power laws in the framework of
weak turbulence theory for surface gravity waves in deep water.
Starting from a random wave field, we let the system evolve numerically 
according to the nonlinear Euler equations for gravity 
waves in infinitely deep water.
In agreement with the theory of Zakharov and Filonenko, 
we find the formation of a power spectrum characterized by a
power law of the form of $|{\bf k}|^{-2.5}$.
\end{abstract}
%
\begin{multicols}{2} 

After the pioneering work by Kolmogorov \cite{k41} on the equilibrium range in 
the spectrum of an homogeneous and isotropic turbulent flow, 
there have been a number of studies on cascade processes in many other
fields of classical physics such as plasma physics, magnetohydrodynamics and ocean waves. 
For surface gravity waves the first seminal theoretical work was done by
O.M. Phillips in 1958, \cite{Phil1}.  Using dimensional arguments, he argued that 
the frequency spectrum in the inertial range was of the form 
$F(\omega)=\alpha g^2 \omega^{-5}$, 
where $\alpha$ was supposed to be an absolute constant and $g$ is gravity.
Even though in the introduction of Phillips' paper it was stated
that ``a necessary condition for the equilibrium range 
over a certain part of the spectrum is the appreciable
 non-linear interactions among these wave-numbers'' (from 
\cite{Phil1}), his arguments were based on the geometrical 
features of the free surface elevation. One of his basic assumptions was that the 
only variable of interest was gravity, while the friction velocity, $u_*$,
was not supposed to be involved in the spectral relation, 
limiting the possibility for a correct dimensional analysis. 

Some years later Zakharov and Filonenko \cite{ZAF} 
established that in infinite water the direct cascade 
should produce a power spectrum of the surface elevation of the 
form $P(|{\bf k}|)\sim |{\bf k}|^{-2.5}$
that corresponds, using the linear dispersion relation in infinite depth, to 
an $\omega^{-4}$  frequency power spectrum: the result was found as
an exact solution of the {\it kinetic wave equation} (see \cite{ZAKH1}). 
The theory developed is known as "weak" or "wave turbulence" and has many
important applications in different fields of physics such as hydrodynamics, 
plasma physics, nonlinear optics, solid state physics, etc. see \cite{FLZ}.
It is called weak turbulence because it deals with resonant interactions
among small-amplitude waves. Thus, contrary to 
fully developed turbulence, it leads to explicit analytical solutions provided
some assumptions are made.
The first experimental support of the theory for surface gravity waves
was made by Toba \cite{TOB} who was completely unaware of the 
paper by Zakharov and Filonenko. He reformulated the Phillips'
equilibrium range law in the
following way: $F(\omega) = \beta g u_* \omega^{-4}$, where $\beta$ 
should now be a universal dimensionless constant. 
After the work by Toba, successive experimental observation of the 
$\omega^{-4}$ law have been made by a number of authors, see for example
\cite{KAH,FOR,DONA,PHIL2}. 

Even though there is a consensus 
on this result, it must be stressed that so far 
the verification of the theory
has never been established from first principles and
moreover the mechanisms that lead to 
the power law $\omega^{-4}$ are not universally recognized:
geometrical aspects related to wave breaking,
 without invoking the nonlinear 
 wave-wave interaction mechanism,
are still retained by many
oceanographers as fundamental for generating an 
$\omega^{-4}$ power law. 
Confirmation of the Zakharov-Filonenko solution to the kinetic equation
has being given through numerical 
simulations of the kinetic wave equation itself 
\cite{RESIO} \cite{PUSH}, solving exactly the so called $S_{nl}$ term. 
Nevertheless, it must be underlined that the kinetic equation is
derived from the primitive equations of motion under a number of hypotheses 
(see for example \cite{ZAKH2}),
therefore it cannot be concluded $a$ $priori$ that power law solutions of the 
kinetic equation  are
also shared by the fully nonlinear wave equations. 

One way to verify weak turbulence theory is
to perform direct numerical simulations 
of the primitive equations of motion.
The numerical confirmation of the theory for gravity waves propagating on a surface 
has not been an easy task 
(for capillary waves see \cite{PUSH1}, for one dimensional wave turbulence 
see \cite{MAI,ZGPD}),
basically because of the intrinsic difficulties of the 
computation of the boundary conditions. Different numerical approaches have been used 
for integrating the fully nonlinear surface gravity waves equations 
(see \cite{YUE} for a review).
The numerical methods based on volume formulations show very interesting results, 
in particular they are capable of modeling in a quite appropriate way  wave breaking.
Unfortunately they have the disadvantage that they require large computational 
resources, and therefore are not suitable for long time numerical simulations. 
For irrotational and inviscid flows boundary formulations are usually preferred:
only the surface is discretized reducing the 
dimension of computation (from three to two). The 
Higher-Order Spectral Methods (HOS), indeed the method used 
in our numerical simulations,  introduced independently by 
West et al. \cite{WEST} and by 
Dommermuth et al. \cite{DOM}, belongs to this second approach
(see also the recent work byTanaka \cite{TAN}).
Very recently three new methods have been proposed 
as very promising for simulating water waves
\cite{SHRIRA,GRU,DYA}. Results using these new approaches on turbulent cascades are
still to be completed. 

In this Letter  we establish numerically, using a HOS method,
that nonlinear interactions
are sufficient for generating  power laws in 
wave spectra; moreover we 
show that the Zakharov-Filonenko theory is 
completely consistent with the primitive equations of motion.
We consider a system of random waves localized 
in wave number space and  we show how
nonlinearities ``adjust'' the spectrum in agreement 
with the Zakharov and Filonenko prediction. 
Numerical work in the case of a forced and dissipated 
system has been attempted by Willemsen \cite{WILL} using 
what sometimes are called the ``Krasitskii equations'' (see
also \cite{ZAKH2}). In order to avoid the effects of external forcing, we considered
the case of a freely decaying wave field. If the simulations, as we will see,
show the formation of a power law
then the
conjecture that this power law is caused by geometrical features related to
forcing and wave breaking must be excluded, since forcing is 
absent and wave breaking cannot be taken into account using the 
numerical method considered. From a physical point of view, 
the freely decaying case corresponds to the evolution of a swell wave field.
A generic wave field is considered at time $t=0$ and it is allowed to evolve
in a natural way using a high order approximation of the Euler equations. 
Since numerical 
computations are limited by the dimension of the 
grid considered, an artificial dissipation is
needed at high wave numbers in order to prevent accumulation 
of energy and a break down of the numerical code. 
The fluid is considered inviscid, irrotational and incompressible. Under these
conditions the velocity potential $\phi(x,y,z,t)$ satisfies
 the Laplace's equation everywhere in the fluid.
The boundary conditions are such that the vertical velocity at the bottom is zero and
on the free surface the kinematic and dynamic boundary conditions are satisfied for the velocity 
potential $\psi(x,y,t)=\phi(x,y,\eta(x,y,t),t)$
 (we assume that fluid is of infinite depth):
\begin{equation}
{\psi_t+g\eta+\frac{1} {2} \bigg[\psi_x^2+\psi_y^2-(\phi_z|_{\eta})^2(1+\eta_x^2+\eta_y^2)\bigg]=0
} \label {Euler1}
\end {equation}
\begin{equation}
{\eta_t+\psi_x\eta_x+\psi_y\eta_y-\phi_z|_{\eta}(1+\eta_x^2+\eta_y^2)=0,
} \label {Euler2}
\end {equation}
The major difficulty for numerical simulations of the system (\ref{Euler1})-(\ref{Euler2})
consists in that we have to compute the derivatives of 
$\phi$ with respect to $z$ on the surface $\eta$. This problem can be overcome 
if we express the velocity potential $\psi(x,y,t)$
as a Taylor expansion around
$z=0$. Inverting asymptotically the expansion 
one can express $\phi_z|_{\eta}$ as an expansion of derivatives of $\psi(x,y,t)$ that 
can then be computed using the Fast Fourier Transform,
simplifying notably the computation. This is nothing other than a different 
way for formulating the HOS method. We underline that this is the same approach that
has originally been used in \cite{ZAKH1} for deriving analytically
the equation that is usually known as the ``Zakharov equation''.
The order of the simulation
can be decided {\it a priori} and depends on how many terms are retained
in the Taylor expansions; in our numerical
simulations we considered the expansion necessary to take into account 
four wave interactions so that we are consistent with the order of
the ``Zakharov equation''. 

A delicate point in our numerical simulations is related
to the dissipation of energy at high
wave numbers. We remark that this dissipation is completely artificial 
since we are dealing with a potential flow. Nevertheless 
we have considered the
dissipation phenomenon of the wave field 
to be similar to the one that takes place in a turbulent flow, i.e. 
that is mathematically expressed by a Laplacian
that operates on the velocity. As is usually done in direct
numerical simulations of box turbulent flows, in order to 
increase the inertial range, we have used a higher order diffusive term. 
More explicitly on the 
right hand side of equation (\ref{Euler1})-(\ref{Euler2}), we have added respectively two
extra terms: $-\nu(-\nabla^2)^n\psi$ and $-\mu(-\nabla^2)^m\eta$, where $\nu$ and 
$\mu$ represent an artificial viscosity coefficient and $\nabla^2$ is the horizontal Laplacian. 
If $n$ and $m$ are greater than 1 the viscosity is known as ``hyperviscosity''.

It has to be noted that, 
at first sight, one would use a very high power of the Laplacian in order 
to increase notably the inertial range, unfortunately very high values of 
$m$ and $n$ could bring about the ``bottleneck effect'' \cite{FALK}, 
i.e. an accumulation of energy at high wave numbers that could
distort the power law expected \cite{BISH}. 
In our numerical simulations
we used $\nu=\mu=3\times 10^4$ and $n=m=8$. These values have 
been selected after some trial
and error during the development of the numerical code: 
because of our limitated number of grid points,
smaller values of $m$ and $n$, 
would obscure almost completely the inertial range.
In our numerical simulations we did not impose any
a dissipation at low wave numbers.

In order to prepare the initial wave field it is reasonable to consider a
directional spectrum $S(|{\bf k}|,\theta )=P(|{\bf k}|)G(\theta )$.
The directional spreading function $G(\theta)$
used here is a cosine-squared function in which only the first lobe (relative
to the dominant wave direction) is considered:  
\begin{equation}
{G(\theta)=\left\{  
\begin{array}{ll}
\ \frac{1} {\sigma} cos^2 \bigg( \frac{\pi} {2 \sigma} \theta \bigg)  
 & if -\sigma \leq \theta \leq \sigma \\
0 & else
\end{array}
\right.}
\label{spreading}
\end{equation}
$\sigma $ is a parameter that
provides a measure of the directional spreading, i.e. as $\sigma \rightarrow 0$,
the waves become increasingly unidirectional. In our numerical
simulations we selected the value of $\sigma=\pi/2$. We tried to avoid the 
complete isotropic case in order to verify if the theory still holds for intermediate 
values of the spreading.
At the same time the selection of a large value
of $\sigma$ was motivated
by the fact that  recently  it has been found \cite{ONO1} that, for
sufficiently narrow angle of spreading, the Benjamin-Feir instability can 
be responsible for the formation of freak waves. As a consequence 
the nonlinear energy transfer could be slightly altered and some corrections
to the prediction could be necessary (this very interesting topic is 
 now under investigation and results will be reported in a different paper). 
$P(|{\bf k}|)$ is chosen to be any localized spectrum. We have performed numerical simulations
with a gaussian function or with a ``chopped JONSWAP'' spectrum (a JONSWAP spectrum with 
amplitudes equal to zero
for frequencies greater than 1.5 times the peak frequency) with random phases. For the case of the
gaussian function, wave numbers lower than a selected threshold have been set to zero in order 
to avoid extremely long large waves. The velocity potential is then computed from the initial wave field 
using the linear theory.
Both gaussian and JONSWAP spectra led to the 
same results in terms of the turbulent cascade. 

\begin{figure} 
\epsfxsize=8.5cm \epsfbox{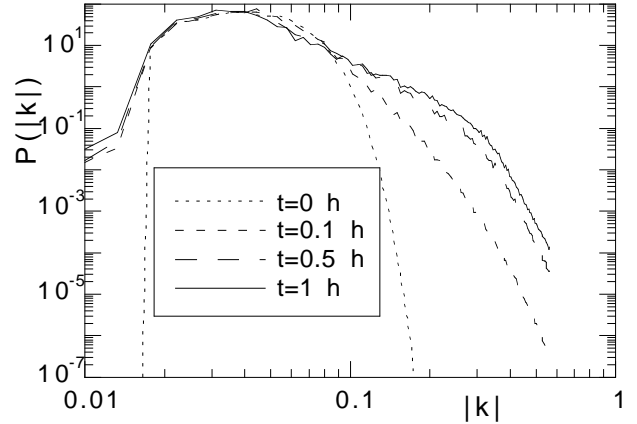}
\caption{Wave spectra at different times.} 
\label{fig spectra1}
\end{figure}

\begin{figure} 
\epsfxsize=8.5cm 
\epsfbox{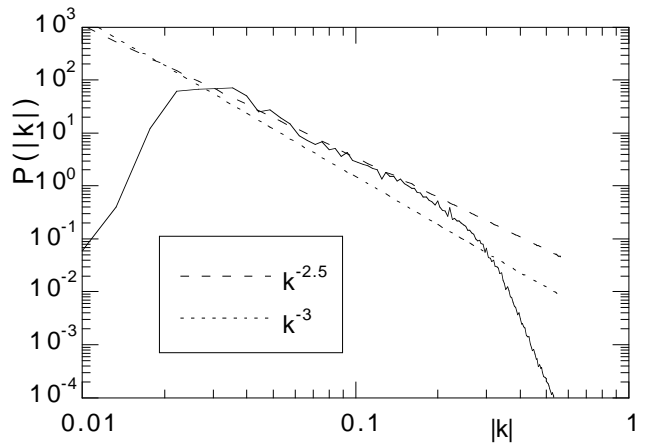}
\caption{Wave spectrum at $t=4$ hours.
A $k^{-2.5}$ (dotted-line) and  a $k^{-3}$ (dashed-line) 
 power law are also plotted.} \label{fig spectra4}
\end{figure}

\begin{figure} 
\epsfxsize=8.5cm \epsfbox{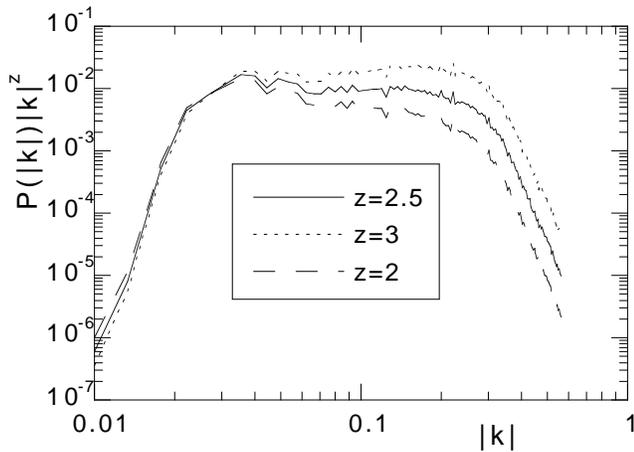}
\caption{Compensated wave spectra for 
different values of the compensation power: $z=2$ (dahsed-line)
$z=2.5$ (solid-line) and $z=3$ (dotted-line).} 
\label{fig spec_comp}
\end{figure}

Our computation is performed in dimensional units;
we have selected the initial spectrum centered at 0.1 Hz, i.e. 
we are considering 10 seconds waves. The initial steepness computed as
$\varepsilon=k_0 H_s/2$ was chosen to be around 0.15 ($H_s$ was computed as
4 times the standard deviation of the wave field). 
The wave field was contained in a square grid (the resolution is $256\times256$)
of length $L=1417.6$ meters.
The time step considered was $1/50$ the dominant frequency,
i.e. $\Delta t=0.2$ seconds. We have performed our numerical simulations
on a 400Mh PC. 
In Fig. \ref{fig spectra1} we show the evolution of the wave
power spectrum for different real times (t=0, 0.1, 0.5, 1 hours).
We see that, as expected, the tail of the
spectrum starts to grow. This process seems to be quite fast: as
is shown in the figure after 
a few dominant wave periods some energy is already injected into high 
wave numbers. The process of adjusting the power law to the 
``correct'' one becomes then very slow, especially for low wave number. 
This could be due to the frozen turbulent phenomenon \cite{PHSH2}, i.e. a condition
in which the energy fluxes towards high wave numbers are reduced because of 
the discretness of the spectrum. 
Moreover decaying numerical simulations are very time consuming
with respect to forced simulations because, as time passes,
energy is lost due to viscosity, thus 
reducing the significant wave height of the wave field and
therefore the steepness. 
 Even though it is not clear from
the Log-Log representation in Fig. \ref{fig spectra1}, there is 
a downshifting of the peak of the spectrum towards
lower wave numbers; as a consequences the steepness subsequently decreases over time.
The time scale of the nonlinear energy 
transfer becomes larger and larger.
In Fig \ref{fig spectra4} we show the power spectrum of the surface 
elevation after 4 hours (the steepnes of the wave field 
is $\varepsilon \simeq 0.07$). In the same plot we show two power laws
$\sim k^{-2.5}$ and $\sim k^{-3}$: the first one seems to better fit the data.
In order to be completely sure that the numerical data are
in agreement with the prediction of Zakharov and Filonenko, we show in 
Fig. \ref{fig spec_comp} compensated spectra with different compensation
powers: $z=2.5$ seems to be the most plausible power law.
Thus there seems to be ample evidence from our numerical simulations that the power law is in 
sufficiently good agreement with the value predicted by the theory.
After the pioneering work by Zakharov and Filonenko the kinetic 
wave theory has developed further, making available 
quantitative predictions for other physical observables such as energy fluxes, 
downshifting of the peak, energy dissipation etc. All these quantities
will be examined and results 
will be reported in future papers. 
Other questions naturally arise from our results: in HOS simulations
the order of the computation depends on how many terms
are retained in the Taylor expansion;
do higher order terms influence the cascade process? 
Our computation has been performed in a freely decaying case;  could 
external forcing (especially if anisotropic) influence the power law?
And more, what would be the influence of the water depth?
These are all questions to be answered in the near future.
%

{\bf Acknowledgements}
This work was supported by  the Office of Naval Research of the
United States of America (T. F. Swean, Jr.) and
by the U.S. Army Engineer Research and Development Center.
Consortium funds and Torino University funds (60 \%) are also acknowledged.
M. O. was also supported by a Research Contract from the Universit\`{a} di
Torino.


\end{multicols} 
\end{document}